\documentstyle[12pt]{article}
\newcommand{\be}{\begin{equation}}
\newcommand{\ee}{\end{equation}}

\newcommand{\bea}{\begin{eqnarray}}
\newcommand{\eea}{\end{eqnarray}}
\newcommand{\unop}{1\!{\rm I}}
\title{Quasi exactly solvable matrix Schr\"odinger operators.}
\author{Yves BRIHAYE\\
Facult\'e des Sciences, Universit\'e de Mons-Hainaut,\\
B-7000 MONS, Belgium.}
\date{\ }
\begin{document}
%%%%%%\begin{titlepage}
\maketitle
\thispagestyle{empty}
\begin{abstract}
\par Two families of quasi exactly solvable $2\times 2$ matrix
Schr\"odinger operators are constructed. The first one is based on a
polynomial matrix potential and depends
 on three parameters. The second
is a one-parameter generalisation of the scalar Lam\'e equation.
The relationship between these operators and QES Hamiltonians
already considered in the literature is pointed out.

\end{abstract}
%%%\vfill
%%%%%%%%\end{titlepage}
%%%\newpage
\section{Introduction}
In a recent paper \cite{sz} a classification of $2\times 2$ matrix
quasi exactly solvable (QES) Schr\"odinger operators in one spatial
dimension is attempted.
This problem was first addressed in \cite{ts} and further 
developped in \cite{bk} and \cite{fgr}.
Here we consider a suitable class of finite dimensional
vector spaces of polynomials in a real variable
and we construct two families of operators preserving 
sub-classes of these vector spaces.
The first family is related to one of the cases treated in \cite{sz};
the second, which is not
considered in \cite{sz},  generalizes
an equation considered in \cite{bk},\cite{bb}.
The two corresponding  QES equations respectively 
constitute ``coupled channel'' generalisations of 
the anharmonic and Lam\'e QES scalar equations.

\par Following the basic idea of QES operators \cite{tur}, \cite{ush} we
consider the finite dimensional vector space of couples of polynomials
of given degree $n$ and $m$ in a real variable $x$. We slightly generalize
this vector space by setting
\be
\label{espace}
{\cal V} = P \left(\begin{array}{l}
{\cal P}(n)\\
{\cal P}(m)
\end{array}\right)
\ee
where ${\cal P}(n)$ denotes the set of real polynomials of degree at
most $n$ in $x$ while $P$ is a fixed invertible $2\times 2$ matrix
operator; $P$ can be interpreted as a change of basis in the vector
space ${\cal P}(n) \oplus {\cal P}(m)$. With such an interpretation
it is reasonnable to choose $P$ of the form
\be
\label{chbase}
P =
\left(\begin{array}{cc}
1 &P_{12}\\
0 &1
\end{array}\right)
\qquad ({\rm{resp.}}\  P =
\left(\begin{array}{cc}
1 &0\\
P_{21} &1
\end{array}\right) )
\ee
for $n\leq m$ (resp. $m\leq n$). In this paper, we limit ourself to
scalar operators $P_{12}$ (or $P_{21})$ of the form
\be
\label{P12}
P_{12} = \kappa_0 {\partial\over{\partial x}} + \kappa_1 + \kappa_2 x{\partial\over
{\partial x}} + \kappa_3 x
\ee
where $\kappa_j$ are constants. The vector space defined
in Eq. (6) of \cite{sz}
can be set in the form (\ref{espace}) with $\kappa_0 = 1$, $\kappa_{1,2,3}=0$.

\section{Polynomial potential}
We consider an operator of the form
\be
H(y) = -{d^2\over{dy^2}} \unop_2 + M_6(y)
\ee
where $M_6(y)$ is a $2\times 2$ hermitian matrix whose entries are
even polynomials of degree at most six in $y$. This operator is a
natural generalisation of the famous QES anharmonic
oscillator \cite{tur,ush} to $2\times 2$ matrix operator. 
After the standard ``gauge transformation''
of $H(y)$ with a factor
\be
\phi(y) = y^{\epsilon} \exp -\lbrace {p_2\over 2} y^4+p_1y^2\rbrace
\ee
and the change of variable $x=y^2$, 
the equivalent operator $\hat H(x)$
can be computed~:
\be
     \hat H(x) = \phi^{-1}(x) H(y) \phi(x)\mid_{y=\sqrt x}
\ee

Then we pose the problem :
what is the most general choice of $M_6$ such that $\hat H(x)$ preserves
a vector space of the form (\ref{espace}),(\ref{chbase}),(\ref{P12})?.

 The following solutions are obtained after straighforward
calculations
(we exclude the case where $M_6(y)$ is diagonal since it corresponds to
a direct sum of two scalar QES operators)~:
\be
\epsilon = 0 \ \ {\rm or} \ \ 1\quad, \quad n=m-2\quad , \quad \kappa_1=\kappa_2=\kappa_3=0 \ .
\ee
The corresponding potentials $M_6$ has the form 
\begin{eqnarray}
M_6(y) &=& \lbrace 4p^2_2y^6+8p_1p_2y^4+(4p^2_1-8mp_2+2(1-2\epsilon)p_2)y^2 \rbrace
\unop_2 \nonumber \\
&+& (8p_2 y^2+4p_1)\sigma_3-8mp_2\kappa_0\sigma_1
\label{m6}
\end{eqnarray}
where $\sigma_1, \sigma_3$ are the Pauli matrices,
$p_2,p_1,\kappa_0$ are free real parameters and $m$ is an integer.
In particular, the non diagonal term is parametrized by an arbitrary
constant which cannot be suppressed because of the $y$-dependent term
proportional to $\sigma_3$. 
If the parameter $\epsilon$ is choosen as an arbitrary real number,
then the potential $M_6$ has a supplementary term of the form
$\epsilon (\epsilon-1)/y^2$.

When the parameters of the case 1 of Ref. \cite{sz} are choosen
so that the potential is a polynomial matrix 
(i.e. $\alpha_2=\alpha_0=0$, $\alpha_1 = 1$, $\beta_0 = 1/2 {\rm \ or \ }
3/2$  in Eq. (34) of \cite{sz}) the potential reduces to the matrix $M_6(y)$ above.
The way of obtaining this result here is slightly different
because the method starts from the natural vector space 
${\cal P}(m) \oplus {\cal P}(n)$.
The more elaborated QES operators
obtained in \cite{sz} can also be produced by our technique
but this is not aimed in this note.

\par Let us also point out that the ``gauge factor $U$'' considered
in \cite{fgr} is limited to be a function of the variable 
$x$ and, therefore,
is not supposed to contain any derivative operator like our operator
$P$ (see (\ref{P12})). This explains that the QES polynomial potential
(\ref{m6}) was not found in \cite{fgr}. With the restriction that $U$
is a function of $x$ only, these authors correctly reach the conclusion
that Hamiltonian preserving a space like ${\cal P} (n) \oplus {\cal P}(m)$
with $|n-m|>1$ are essentially diagonal, in contrast with the present
operator related to the case $n-m=2$.

\section{Application to N-body hamiltonians}
By using the idea of the previous section, a matrix version of the QES 
many-body problem of Ref. \cite{mrt} can be constructed.
Let us consider the Calogero Hamiltonian \cite{cal} (we note it $H_{cal}$) 
supplemented by a matrix-valued potential  $V^*$~:
\be
\label{calo}
    H = H_{cal} + V^{*} = \frac{1}{2} \sum_{j=1}^N 
 [- \frac{\partial^2}{\partial x_j^2} + x_j^2]
   + \sum_{j<i} \frac{\nu (\nu-1)}{(x_j-x_i)^2} + V^{*}
\ee
Along with \cite{mrt}
we assume $V^*$ to depend only on the variable $\tau$
\be
    \tau \equiv \sum_{j<i}^N(x_j-Y)(x_i-Y) \ \ \ , \ \ \ 
     Y \equiv \sum_{j=1}^N x_j \ \ ,
\ee
and we look for eigenfunctions of the hamiltonian (\ref{calo})
 of the form
\be
     \Psi(x) = \psi_0(x) \  
\ \tau^{\epsilon} \exp -\lbrace {p_2\over 2} \tau^4+p_1 \tau^2\rbrace 
\ \phi (\tau)
\ee
where $\psi_0$ denotes the ground state of the 
standard Calogero system~:
\be 
        \psi_0(x) =  \Bigl[ \prod_{i<j} \vert x_i - x_j \vert)
\Bigr]^{\nu} \exp (- X^2/2)  \ \ \ , \ \ 
X^2 \equiv \sum_{j=1}^N x_j^2     
\ee 
while $\phi(\tau)$ represents a couple of polynomials in $\tau$.

After a standard algebra,  
the operator acting on $\phi(\tau)$ can be isolated~:
\be
   h \equiv   \tau \frac{\partial^2}{\partial \tau^2} 
       + (4 \tau + 2 b)\frac{\partial}{\partial \tau} + V^*
\ee
and
it can be shown that this operator preserves the space
(\ref{espace}) (again with $n=m-2$, $\kappa_1=\kappa_2=\kappa_3=0$)
provided  $V^*$ is of the form 
\be
\label{poten}
      V^*(\tau) =  - p_2^2 \tau^3 + 2p_2(1-p_1)\tau^2 + (a-2p_2 \sigma_3)\tau
      + (1-p_1)\sigma_3 + \frac{\gamma}{\tau} + 2 m \kappa_0 \sigma_1
\ee
with the definitions
\be
    a \equiv p_1(2-p_1) + p_2(2m+3\epsilon-1+b) \ , \ 
    b\equiv \frac{1}{2} (1+\nu N)(N-1) \ , \ \gamma \equiv 2\epsilon(\epsilon-1+b) \ .
\ee
As a consequence, (\ref{calo}),(\ref{poten}) constitutes a QES matrix
extension (labelled by the parameters $p_1, p_2, \epsilon$)
of the exactly solvable Calogero hamiltonian.

\section{Lam\'e type potential.}
\par As a second example, we consider the family of operators
\be
\label{lame}
H(z) = -{d^2\over{dz^2}} +
\left[\begin{array}{cc}
Ak^2{\rm{sn}}^2+\delta (1+k^2)/2 &2\theta k {\rm{cn}}\ {\rm{dn}}\\
2\theta k {\rm{cn}}\ {\rm{dn}} &Ck^2{\rm{sn}}^2-\delta (1+k^2)/2
\end{array}\right]
\ee
where $A,C,\delta, \theta$ are constants while ${\rm{sn}}, {\rm{cn}},{\rm{dn}}$ respectively abbreviate
the Jacobi elliptic functions of argument $z$
and modulus $k$ \cite{arscott}
\be
{\rm{sn}}(z,k) \quad , \quad {\rm{cn}} (z,k)\quad , \quad {\rm{dn}}(z,k)
\ \ .
\ee
These functions are periodic with period $4K(k),
4K(k), 2K(k)$ respectively ($K(k)$
is the complete elliptic integral of the first type). The above
hamiltonian is therefore to be considered on the Hilbert space of
periodic functions on $[0,4K(k)]$. 
For completeness, we mention the properties of the Jacobi
functions  which are needed in the calculations 
\be
{\rm{cn}}^2+{\rm{sn}}^2=1\quad , \quad {\rm{dn}}^2+k^2{\rm{sn}}^2=1
\ee
\be
{d\over{dz}}{\rm{sn}} = {\rm{cn}}\  {\rm{dn}}\quad ,
\quad {d\over{dz}} {\rm{cn}} = -{\rm{sn}}\  {\rm{dn}}\quad
, \quad {d\over{dz}} {\rm{dn}} = -k^2 {\rm{sn}}\  {\rm{cn}}
\ee

The relevant change of variable which 
eliminates the transcendental functions sn, cn, dn from (\ref{lame})
in favor of algebraic expressions is (for $k$ fixed)
\be
x={\rm{sn}}^2(z,k)
\ee
In particular the second derivative term in (\ref{lame})
becomes
\be
      {d^2 \over dz^2} = 
4x(1-x)(1-k^2x) {d^2 \over d^2 x} 
+ 2(3 k^2 x^2 - 2 (1+ k^2)x + 1){d \over d x} 
\ee
Several possibilities of extracting prefactors then lead 
to equivalent forms of (\ref{lame}), say $\hat H(x)$,
which are matrix operators build with the derivative $d/dx$
and polynomial coefficients in $x$.
The requirement that $\hat H(x)$ preserves a
space of the form (\ref{espace}) leads to two possible sets of
values for
$A,C,\theta$ (we do not consider the case $\theta=0$ since it 
corresponds to
two decoupled scalar Lam\'e equations).

\smallbreak
\noindent {\sc{Case 1}}
\smallbreak

\noindent $A=4m^2+6m+3-\delta$\\
$C=4m^2+6m+3+\delta$\\
$\theta = {1\over 2} [(4m+3)^2-\delta^2]^{1\over 2}$\\
The parameter $\delta$ remains free, and also $k$ which fixes the period
of the potential.
Four invariant spaces are available. In order to present them we
conveniently define
\be
R_1 = {4m+3-\delta\over{4m+3+\delta}},
\ee
We have then
\begin{eqnarray}
{\cal V}_1 &=&
\left(\begin{array}{cc}
1 &0\\
0 &{\rm{cn}}\ {\rm{dn}}
\end{array}\right)
\left(\begin{array}{cc}
1 &\kappa x\\
0 &1
\end{array}\right)
\left(\begin{array}{c}
{\cal P}(m)\\
{\cal P}(m)
\end{array}\right)
\quad , \quad \kappa^2=k^2R_1\\
{\cal V}_2 &=&
\left(\begin{array}{cc}
{\rm{cn}}\ {\rm{dn}} &0\\
0 &1
\end{array}\right)
\left(\begin{array}{cc}
1 &0\\
\kappa x &1
\end{array}\right)
\left(\begin{array}{c}
{\cal P}(m)\\
{\cal P}(m)
\end{array}\right)
\quad , \quad \kappa^2=k^2/R_1\\
{\cal V}_3 &=&
\left(\begin{array}{cc}
{\rm{sn}}\ {\rm{cn}} &0\\
0 &{\rm{sn}}\ {\rm{dn}}
\end{array}\right)
\left(\begin{array}{cc}
1 &\kappa\\
0 &1
\end{array}\right)
\left(\begin{array}{c}
{\cal P}(m-1)\\
{\cal P}(m)
\end{array}\right)
\quad , \quad \kappa^2=k^2R_1\\
{\cal V}_4 &=&
\left(\begin{array}{cc}
{\rm{sn}}\ {\rm{dn}} &0\\
0 &{\rm{sn}}\ {\rm{cn}}
\end{array}\right)
\left(\begin{array}{cc}
1 &\kappa\\
0 &1
\end{array}\right)
\left(\begin{array}{c}
{\cal P}(m-1)\\
{\cal P}(m)
\end{array}\right)
\quad , \quad \kappa^2=R_1/k^2
\end{eqnarray}

\smallbreak
\noindent{\sc{Case 2}}
\smallbreak

\noindent
$A=4m^2+2m+1-\delta$\\
$C=4m^2+2m+1+\delta$\\
$\theta={1\over 2} [(4m+1)^2-\delta^2]^{1\over 2}$\\
The associated invariant vector spaces read, defining $R_2=(4m+1-\delta)/(4m
+1+\delta)$,

\begin{eqnarray}
{\cal V}_5 &=&
\left(\begin{array}{cc}
{\rm{cn}} &0\\
0 &{\rm{dn}}
\end{array}\right)
\left(\begin{array}{cc}
1 &\kappa\\
0 &1
\end{array}\right)
\left(\begin{array}{c}
{\cal P}(m-1)\\
{\cal P}(m)
\end{array}\right)
\quad , \quad \kappa^2=k^2R_2\\
{\cal V}_6 &=&
\left(\begin{array}{cc}
{\rm{dn}} &0\\
0 &{\rm{cn}}
\end{array}\right)
\left(\begin{array}{cc}
1 &\kappa \\
0 &1
\end{array}\right)
\left(\begin{array}{c}
{\cal P}(m-1)\\
{\cal P}(m)
\end{array}\right)
\quad , \quad \kappa^2=R_2/k^2\\
{\cal V}_7 &=&
\left(\begin{array}{cc}
{\rm{sn}} &0\\
0 &{\rm{sn}}\ {\rm{cn}}\ {\rm{dn}}
\end{array}\right)
\left(\begin{array}{cc}
1 &\kappa x\\
0 &1
\end{array}\right)
\left(\begin{array}{c}
{\cal P}(m-1)\\
{\cal P}(m-1)
\end{array}\right)
\quad , \quad \kappa^2=k^2R_2\\
{\cal V}_8 &=&
\left(\begin{array}{cc}
{\rm{sn}}\ {\rm{cn}}\ {\rm{dn}} &0\\
0 &{\rm{sn}}
\end{array}\right)
\left(\begin{array}{cc}
1 &0\\
\kappa x &1
\end{array}\right)
\left(\begin{array}{c}
{\cal P}(m-1)\\
{\cal P}(m-1)
\end{array}\right)
\quad , \quad \kappa^2=k^2/R_2
\end{eqnarray}
The operator (\ref{lame}) was studied in \cite{bk},\cite{bb} for $\delta=1$.
For this particular value of $\delta$, the corresponding eigenvalue equation
$H \psi = \omega^2 \psi$ determines the normal modes 
of the sphaleron classical solution \cite{km} in
the Abelian Higgs model in 1+1 dimension. 
It therefore plays a crucial role in the understanding of the instabilities 
of the sphaleron in this model.

The above results demonstrate that the remarkable
algebraic properties of the Lam\'e equation \cite{arscott} also hold
for the operator (\ref{lame}), irrespectively of the value of $\delta$.
The associated eigenvalue equation  therefore constitutes a (one parameter)
$2\times 2$ matrix equation analog of
the scalar Lam\'e equation. 

\section{Generalization}
The kind of operators presented in Sect.~3 can be generalized to matrix 
potentials of the form
\be
\label{lameg}
H(z) = -{d^2\over{dz^2}} +
\left[\begin{array}{cc}
V_1({\rm{sn}}^2) &\theta\ {\rm{sn}}^{\alpha_1}\ {\rm{cn}}^{\alpha_2}\ {\rm{dn}}^{\alpha_3}\\
\theta\ {\rm{sn}}^{\alpha_1}\ {\rm{cn}}^{\alpha_2}\ {\rm{dn}}^{\alpha_3}          &V_2({\rm{sn}}^2)
\end{array}\right]
\ee
where $V_1, V_2$ are polynomials, $\theta$ is a constant and $\alpha_j$ are 
non-negative integers.

The similarity transformation
\be
    \hat H(x) = U^{-1}(z) H(z) U(z) \ \ , \ \ 
     U(z) = {\rm diag} ({\rm{sn}}^{\beta_1}\ {\rm{cn}}^{\beta_2}\ {\rm{dn}}^{\beta_3},
{\rm{sn}}^{\gamma_1}\ {\rm{cn}}^{\gamma_2}\ {\rm{dn}}^{\gamma_3})
\ee 
sets the operator (\ref{lameg}) into a form with polynomial coefficients
in the variable $x = {\rm sn}^2$ provided
\begin{itemize}
\item $\beta_j, \gamma_j = 0 \ {\rm or} \ 1$ \ \ , \ \ $j=1,2,3$
\item $\alpha_j \pm (\beta_j - \gamma_j) =$ non-negative even integer \ \ , \ \ $j=1,2,3$.
\end{itemize}

After making a choice of $\alpha_j, \beta_j, \gamma_j$ satisfying the above
conditions, the possible forms of $V_1, V_2$ and of $P,m,n$ in Eq. (\ref{espace})
have to be determined in order for $H(z)$ to be QES. 

Taking $k=0$, the standard trigonometric functions are recovered~:
\be
   \rm{sn}(z,0) = \sin{z} \ \ , \ \ \rm{cn}(z,0) = \cos{z} \ \ , \ \ 
   \rm{dn}(z,0) = 1 \ .
\ee
The  periodic potential below, which is exactly solvable \cite{bt},
 furnishes a particular example of this type
\be
     V(z) \div   \left[\begin{array}{cc}
{\rm{cos}}^2(z) & {\rm{cos}}(z) \ {\rm{sin}}(z)  \\
  {\rm{cos}}(z)\ {\rm{sin}}(z)         &{\rm{sin}}^2(z)
\end{array}\right] \ \ .
\ee
It determines the normal modes about 
some static solutions of the Goldstone model in 1+1 
dimensions \cite{bt}.

\section{Concluding remarks}
The examples of operators presented above give  evidences
of the difficulty to classify the coupled-channel
(or matrix) QES Schrodinger equations.
The way of constructing the QES  potential
$M_6$ in Sect. 2 further provides a clear link between the
approaches \cite{sz} and \cite{fgr} to this mathematical problem;
we hope that this note will  motivate further investigations
of it.

\end{document}